\documentclass[11pt,a4paper]{article}
\usepackage{jcappub}
\usepackage{float}
\usepackage{indentfirst}
\newcommand{\be}{\begin{equation}}
\newcommand{\ee}{\end{equation}}

\newcommand{\ba}{\begin{eqnarray}}
\newcommand{\ea}{\end{eqnarray}}

\usepackage{float}

\newcommand{\bege}{\begin{equation}}
\newcommand{\bpartial}{\mathop{\partial\kern -4pt\raisebox{.8pt}{$|$}}}
\newcommand{\enge}{\end{equation}}
\newcommand{\beq}{\begin{eqnarray}}
\newcommand{\benu}{\begin{enumerate}}

\newcommand{\enu}{\end{enumerate}}
\newcommand{\eeq}{\end{eqnarray}}

\newcommand{\noi}{\noindent}

\begin{document}

\title{{Gravity with extra dimensions and dark matter interpretation:\\ A straightforward approach}}

\author[a]{C. H. Coimbra-Ara\'ujo}
\author[b]{Rold\~ao da Rocha}
\affiliation[a]{Campus Palotina, Universidade Federal do Paran\'a,
UFPR, 85950-000, Palotina, PR, Brazil.} \affiliation[b]{Centro de
Matem\'atica, Computa\c c\~ao e Cogni\c c\~ao, Universidade Federal
do ABC,  09210-170, Santo Andr\'e, SP, Brazil}
\emailAdd{carlos.coimbra@ufpr.br}
\emailAdd{roldao.rocha@ufabc.edu.br}

\begin{abstract}{
Any connection between dark matter and extra dimensions can be cognizably evinced from the associated effective energy-momentum tensor. In order to investigate and test such relationship, a higher dimensional spacetime endowed with a factorizable general metric is regarded to derive a general expression for the stress tensor -- from the Einstein-Hilbert action -- and to elicit the effective gravitational potential. A particular construction for the case of six dimensions is provided, and it is forthwith revealed that the missing mass phenomenon may be explained, irrespective of the dark matter existence. Moreover, the  existence of extra dimensions in the universe accrues the possibility of a straightforward mechanism for such explanation.
A configuration which density profile coincides with the Newtonian
potential for spiral galaxies is constructed, from a 4-dimensional isotropic
metric plus extra-dimensional components. A Miyamoto-Nagai \emph{ansatz} is
used to solve Einstein equations. The stable rotation curves associated to such
system are computed, in full compliance to the observational data, without fitting techniques. The density profiles are
reconstructed and compared to that ones obtained from the Newtonian
potential. 
   }
\end{abstract}

\flushbottom
\maketitle

\section{Introduction}

General relativity (GR) provides solutions to the Einstein equations
that currently offer important support to understand, for instance,
black holes and cosmology. Mainly in  the last two decades, such
formalism was extended for extra-dimensional spacetimes.
There are many motivations to consider such approach: from the
Kaluza-Klein  -- and string theory as well -- idea of unified fields,
to the recent effort to deal with the hierarchy problem from the
idea of diluting gravity in submillimetric extra dimensions
\cite{randall}. Another motivation -- indeed more fundamental for GR
-- is that higher dimensions may play a fundamental role to unravel
profound interpretations concerning the nature of spacetime. On this
behalf, a series of works deal with a $(4+n)$-dimensional metric
splitting, in order to find solutions for the Einstein equations, in
both static or stationary spherical/axisymmetric geometries. The
static, spherically symmetric solutions of Kaluza-Klein theory were
calculated independently by several authors \cite{leutwyler:60}. For
instance, in \cite{dobiasch:82} and \cite{pollard:83} some
stationary spherically symmetric solutions on Kaluza-Klein theory
are provided, as well as in 5-dimensional GR
\cite{chodos:82}. Furthermore, in \cite{gibbons:86} some
asymptotically flat Kaluza-Klein solutions, corresponding to regular
black holes in four dimensions, are regarded, and prominent features
concerning thermal evaporation are presented. Such formalism is
generalized in a $(4 + n)$-dimensional scenario \cite{gibbons:84}. It is accomplished in
the so called abelian Kaluza-Klein theory --- possessing the U(1)$^n$ as its
internal isometry group --- which lower dimensional case was
investigated in \cite{cvetic:95}. The rotating dyonic
black holes of Kaluza-Klein theory were investigated
\cite{rasheed:95} withal, and the explicit
connection between such models and extra-dimensional theories is explored in \cite{wesson}.

Therewith, one of the most outstanding non compact extra dimensions
usage is the Randall-Sundrum (RSI) model, wherein the so called
braneworld is embedded in an AdS$_5$ space \cite{randall}. Moreover,
some large extra dimensions formalisms were investigated in a
(4+1+1)-dimensional setting. The Einstein equations with negative
bulk cosmological constant and a 3-brane, with appropriately tuned
energy momentum tensor, were solved \cite{gregor}. In the case of
two extra dimensions, which shall be considered as a particular case
of our formalism, some aspects were approached at \cite{rubakov}.

The pivotal physical motivation in our formulation is that gravity
with extra dimensions can possibly shed new light on the so called
``dark matter'' puzzle \cite{coimbra1,coimbra2,kahil,coimbra3}. The
idea is not a novelty, since some theories, as universal extra
dimensions \cite{acd}, and some variants of Kaluza-Klein theories as
well, present particles corresponding to extra-dimensional
excitation modes -- representing a dark matter candidate (for a
review see \cite{profumo}). Here the approach is different, in the
sense that it is argued whether the
 existence of higher dimensional spacetimes suffices to explain the missing mass problem ({\it aka} ``dark
matter''), without any specific dark particle.

In order to precisely ascertain what this assumption infuses in the dark matter problem, we delve into a generalization where
our universe has $D = 4 + n$ dimensions. Moreover, at a first glance
nothing precludes the matter to live -- or leak as well --  in the $n$
extra dimensions (see e.g. at the level of particle physics, the UED
proposition \cite{acd}). In what follows the formalism is developed {\it
a la} Kaluza-Klein

Accordingly from such scenario, an Einstein--Hilbert
gravitational action
\begin{equation}\label{action}
S=\int \mathrm{d}^4x\,\mathrm{d}^n y
\sqrt{-^{(4+n)}g}~\left(^{(4+n)}R+\mathcal{L}_M\right),
\end{equation} where $\mathcal{L}_M$ denotes the matter Lagrangian, leads to the field equations \bege\label{einstein}
~^{(4+n)}G_{AB}= ~^{(4+n)}T_{AB}, \enge \noi where $A,B = 0, 1, \ldots,
4+n$, and $y$ denotes the extra dimensions. Hereon  the tensor components
$~^{(4+n)}G_{AB}$ and $~^{(4+n)}T_{AB}$ are respectively denoted by
$G_{AB}$ and $T_{AB}$ (the same for the curvature tensor and scalar,
and for the metric components as well). No modification is considered for
the gravitational constant $G$.

Given such approach, in Section \ref{sec:splitting} the spacetime
metric is split, in order to express the  4-dimensional stress
tensor. It infuses some interpretations about the extra terms,
appearing due to the inclusion of new dimensions (Section
\ref{sec:discussion}). In Section \ref{sec:motion} the equations of
motion are developed and in Section \ref{sec:visible} we calculate
-- by linearized gravity -- the Newtonian limit and a modified
Poisson equation for the theory. The Section \ref{sec:extra} is
dedicated to show how one can interpret dark matter as a prominent
pure effect, coming from the spacetime extra dimensions. We provide
particular cases where  two extra dimensions are regarded. Stable
galaxy rotation curves are presented together with their associated
analytical expression when the system is axisymmetric in Section
\ref{sec:solutions}. Finally in Section \ref{sec:galaxies} some
results with real galaxies are presented and analyzed what, together
with the results in Section \ref{sec:solutions}, put our formalism
in compliance with observational data.

A companion paper shows a phenomenological example \cite{coimbra3},
as well as other semi-phenomenological ones provided by
\cite{coimbra2}. The assumptions here are straightforward, and
quantum field theory is not explicitly taken into account. GR is the
theoretical limit concerning the present work. In this article,
unless  explicitly mentioned, $c=1=8\pi G$, and the signature
adopted for the $4$-dimensional part of the metric is
diag$(-,+,+,+)$. In addition, it is considered {\it ab initio} that
the extra-dimensional part is constituted by spacelike coordinates.

\section{Splitting the metric}\label{sec:splitting}

Our universe is supposed to have $3+1+n$ dimensions. Initially,
the extra dimensions compactification is kept apart. The most
general metric for such universe is given by \bege\label{metric1}
g_{AB}=\begin{pmatrix}
g_{\alpha\beta}&g_{\alpha b}\\
g_{a\beta}&g_{ab}
\end{pmatrix},
\enge \noi where $\alpha,\beta=0,\ldots,3$ and $a,b=4,\ldots, n$.
Furthermore we consider the convention to make the metric as a
function of $3+1$ coordinates only: $g_{AB} = g_{AB}(x^\alpha)$.
This metric components $g_{AB}$ contain the $3+1$ universe metric
terms $g_{\alpha\beta}$ and the extra dimensional terms $g_{ab}$, as
well as the crossed components. Eq. (\ref{metric1}) can be rewritten
for convenience as \bege g_{AB}=
g_{\alpha\beta}\delta^\alpha_A\delta^\beta_B +
g_{ab}\delta^a_A\delta^b_B+g_{\alpha
b}\delta^\alpha_A\delta^b_B+g_{a\beta}\delta^a_{A}\delta^\beta_B,\nonumber\enge
\noi where $\delta^i_j$ are the Kronecker symbols. The derivatives
for such metric components are given by \beq\label{deriv1}
g_{AB,C}=&&g_{\alpha\beta,\gamma}\delta^\alpha_A\delta^\beta_B\delta^\gamma_C
+ g_{ab,\gamma}\delta^a_A\delta^b_B\delta^\gamma_C+g_{\alpha
b,\gamma}\delta^\alpha_A\delta^b_B\delta^\gamma_C+g_{a\beta,\gamma}\delta^a_{A}\delta^\beta_B\delta^\gamma_C.\eeq
The case $g_{AB}= g_{\alpha\beta}\delta^\alpha_A\delta^\beta_B+
g_{ab}\delta^a_A\delta^b_B$ is considered here, motivated by
formalisms where $g^a_\alpha=0$ --- namely Randall-Sundrum warped
metrics \cite{randall} and  other physically relevant cases
\cite{wesson,coimbra1,etc,etc2,etc3}. In particular, a comprehensive
program on observations and measurements on braneworld effects in
astrophysics and particle physics is provided in \cite{etcmeio}.
Furthermore, some generalizations regarding variable tension
braneworld models and cognizable  physical effects were presented
\cite{etc3}.

The inverse metric is
written as
\bege \label{metric_inv} g^{AB}=
g^{\alpha\beta}\delta_\alpha^A\delta_\beta^B+
g^{ab}\delta_a^A\delta_b^B,\enge \noi and the derivatives are
straightforwardly provided by Eq. (\ref{deriv1})
\bege  \label{deriv2}
g_{AB,C}=g_{\alpha\beta,\gamma}\delta^\alpha_A\delta^\beta_B\delta^\gamma_C
+ g_{ab,\gamma}\delta^a_A\delta^b_B\delta^\gamma_C. \enge
\noi Assuming that the spacetime has a connection presenting \emph{no} torsion, one yields the
following Christoffel symbols $\left\{_{BC}^{~A} \right\} =
\frac{1}{2}g^{AM}(g_{BM,C}+g_{CM,B}-g_{BC,M}).$
\noi Splitting this last expression by Eqs. (\ref{metric_inv}) and
(\ref{deriv2}) it reads \bege\label{chris_split} \left\{_{BC}^{~A}
\right\}=
\left\{^{~\alpha}_{\beta\gamma}\right\}\delta^A_\alpha\delta^\beta_B\delta^\gamma_C+
\frac{1}{2}\left[g^{am}(g_{bm,\gamma}\delta^A_a\delta^b_B\delta^\gamma_C+g_{cm,\beta}\delta^A_a\delta^\beta_B\delta^c_C)-g^{\alpha\mu}g_{bc,\mu}\delta^A_\alpha\delta^b_B\delta^c_C\right].
\enge
\noi The Ricci tensor components are well know to be written as
\bege R_{AB} = \partial_M\left\{^{~M}_{A B}\right\}
-\partial_B\left\{^{~M}_{A M}\right\} +
\left\{^{~N}_{AB}\right\}\left\{^{~M}_{NM}\right\} -
\left\{^{~N}_{AM}\right\}\left\{^{~M}_{NB}\right\}.\nonumber\enge
\noi Taking into account that the terms of the
metric depends solely on $x^\alpha$, the equation above reads \bege \label{ricci_split}
R_{AB}=R_{\alpha\beta}\delta^\alpha_A\delta^\beta_B +
R_{ab}\delta^a_A\delta^b_B. \enge \noi  The splitting (\ref{metric_inv}) and the
derivatives (\ref{deriv2}) of the metric in the Christoffel
symbols (\ref{chris_split}) lead to the following expressions for
Eq. (\ref{ricci_split})
\beq
\label{ricci_explicit}  \hspace{-1.3cm}R_{\alpha\beta} &=& \mathcal{R}_{\alpha\beta}
- \frac{1}{2}\left[g^{mc}_{\;\;\;,\beta}g_{mc,\alpha}+
g^{mc}g_{mc,\alpha\beta}+\frac{1}{2}g^{nc}g_{mc,\alpha}g^{me}g_{ne,\beta}\right],\\
\label{ricci_explicit_2} \hspace{-2cm}R_{ab} &=&
-\frac{1}{2}\left[g^{\mu\gamma}_{\;\;\;,\mu}g_{ab,\gamma}+g^{\mu\gamma}g_{ab,\gamma\mu}-\frac{1}{2}\left(g^{nc}g_{ac,\mu}g^{\mu\gamma}g_{nb,\gamma}
g^{\nu\gamma}g_{am,\gamma}g^{mc}g_{bc,\nu}-g^{\nu\gamma}g_{ab,\nu}g^{mc}g_{mc,\gamma}\right)\right]
\eeq
\noi where $\mathcal{R}_{\alpha\beta} =
\partial_\mu\{^{~\mu}_{\alpha \beta}\} - \partial_\beta\left\{^{~\mu}_{\alpha \mu}\right\} +
\{^{~\nu}_{\alpha\beta}\}\left\{^{~\mu}_{\nu\mu}\right\} -
\left\{^{~\nu}_{\alpha\nu}\right\}\{^{~\mu}_{\nu\beta}\}$ is the conventional (3+1) Ricci
tensor.

\section{The
energy-momentum tensor}\label{sec:discussion}
Consider now the Einstein tensor components
$G_{AB}=R_{AB}-(1/2)Rg_{AB}$, where $R=g^{MN}R_{MN}$ is the scalar curvature. For a
stress tensor written as
$T_{AB}=T_{\alpha\beta}\delta^\alpha_A\delta^\beta_B +
T_{ab}\delta^a_A\delta^b_B$, one can follow the same arguments accomplished for
$R_{AB}$. Introducing the results obtained in the previous Section in Eq. (\ref{einstein}), the split components of the stress tensor is given by:
\beq \label{stress} T_{\alpha\beta} &=& R_{\alpha\beta} -
\frac{1}{2}(g^{MN}R_{MN})g_{\alpha\beta},\qquad\qquad
T_{ab}= R_{ab} - \frac{1}{2}(g^{MN}R_{MN})g_{ab}.\eeq \noi The
tensor $T_{\alpha\beta}$ represents the energy/pressure content in
the $(3+1)$-dimensional landscape. This is the part that is clearly of
paramount interest. Regardless of $T_{ab}\ne 0$, here it is assumed that only the
$T_{\alpha\beta}$ visible part can bring some light to
observational issues. From the expression developed for
$R_{\alpha\beta}$, Eq. (\ref{ricci_explicit}), the influence of
extra dimensions is evident. From (\ref{ricci_explicit}),
(\ref{ricci_explicit_2}), and (\ref{stress}) it yields
 \bege \label{stress2} T_{\alpha\beta} = \mathcal{T}_{\alpha\beta} + \mathfrak{T}_{\alpha\beta}, \enge
 \noi where $\mathcal{T}_{\alpha\beta}$ represents the
 $(3+1)$-dimensional influence, and $\mathfrak{T}_{\alpha\beta}$ denotes the correction
elicited uniquely from extra dimensions, where explicitly
\bege \label{talfabeta} \mathcal{T}_{\alpha\beta} =
\mathcal{R}_{\alpha\beta} -
\frac{1}{2}(g^{\mu\nu}\mathcal{R}_{\mu\nu})g_{\alpha\beta},\enge
\bege \label{stress_extra} \mathfrak{T}_{\alpha\beta}=
-\frac{1}{2}\left\{\mathfrak{R}_{\alpha\beta} -
\frac{1}{2}(g^{\mu\nu}\mathfrak{R}_{\mu\nu})g_{\alpha\beta}
+(g^{mn}R_{mn})g_{\alpha\beta}\right\}.\enge
\noi Note that the stress $\mathcal{T}_{\alpha\beta}$ has the
conventional form for the Einstein tensor
$\mathcal{G}_{\alpha\beta}$. The tensor $\mathfrak{R}_{\alpha\beta}$
represents a ``curvature'' term evinced from the presence of
extra dimensions, and is given by
\bege \mathfrak{R}_{\alpha\beta} =
g^{mc}_{\;\;\;,\beta}g_{mc,\alpha}+
g^{mc}g_{mc,\alpha\beta}+\frac{1}{2}g^{nc}g_{mc,\alpha}g^{me}g_{ne,\beta}.
\enge
In addition, from the action (\ref{action}) the stress tensor can be derived
from the conventional definition $T_{AB} := -2\frac{\delta
\mathcal{L}_M}{\delta g^{AB}} + g_{AB}\mathcal{L}_M$. In this new
approach, it is clear that the $(3+1)$-dimensional part $T_{\alpha\beta}$ is
proportional to the Lagrangian multiplied by $g_{\alpha\beta}$ and
some quantity from the variation $\delta/\delta g^{AB}$.
\section{An expression for the equations of
motion}\label{sec:motion}
A general Lagrangian for the gravitating test particles in a spacetime with extra dimensions,
with metric elements $g_{AB}=
g_{\alpha\beta}\delta^\alpha_A\delta^\beta_B+
g_{ab}\delta^a_A\delta^b_B$, can be written as
\begin{eqnarray}\label{lagrangian}
L &=& (g_{AB}\dot{x}^A\dot{x}^B)^{1/2}= (g_{\alpha\beta}\dot{x}^\alpha\dot{x}^\beta + g_{a
b}\dot{x}^a\dot{x}^b)^{1/2},
\end{eqnarray}
\noindent where $\dot{x}^A = \mathrm{d}x^A/\mathrm{d}s$. The motion
equations come from the Euler-Lagrange expression
$
\frac{\mathrm{d}}{\mathrm{d}s}\left(\frac{\partial L}{\partial
\dot{x}^C}\right)-\frac{\partial L}{\partial x^C} = 0.
$
 As $\partial_A = \partial_\alpha\delta^\alpha_A +
\partial_a\delta^a_A$ and $g_{ab} = g_{ab}(x^\alpha)$ it follows that
\begin{eqnarray}
\frac{\partial L}{\partial x^C}&=& \frac{\partial L}{\partial
x^\gamma}\delta_C^\gamma + \frac{\partial L}{\partial
x^c}\delta_C^c,\nonumber\\
\frac{\partial L}{\partial x^\gamma} &=&
\frac{1}{2}L^{-1}(g_{\alpha\beta,\gamma}\dot{x}^\alpha\dot{x}^\beta
+ g_{ab,\gamma}\dot{x}^a\dot{x}^b),\nonumber
\end{eqnarray}
\noi and
$\frac{\partial L}{\partial x^c} = 0.$ It immediately yields
\begin{equation}
\frac{\partial L}{\partial x^C} =
\frac{1}{2}L^{-1}(g_{\alpha\beta,\gamma}\dot{x}^\alpha\dot{x}^\beta
+ g_{ab,\gamma}\dot{x}^a\dot{x}^b).\nonumber
\end{equation}
\noindent Likewise, the term
$\frac{\mathrm{d}}{\mathrm{d}s}\left(\frac{\partial L}{\partial
\dot{x}^C}\right)$ can be developed:
\beq
\frac{\partial L}{\partial \dot{x}^\gamma} =
\frac{1}{2}L^{-1}(g_{\gamma\beta}\dot{x}^\beta +
g_{\alpha\gamma}\dot{x}^\alpha) &=& L^{-1}g_{\mu\gamma}\dot{x}^\mu,\nonumber\\
\frac{\partial L}{\partial \dot{x}^c} =
\frac{1}{2}L^{-1}(g_{cb}\dot{x}^b + g_{ac}\dot{x}^a) &=&
L^{-1}g_{mc}\dot{x}^m.\nonumber
\eeq
\noindent Now
\begin{equation}
\frac{\mathrm{d}}{\mathrm{d}s}\left(\frac{\partial L}{\partial
\dot{x}^\gamma}\right)=L^{-1}\left[\left(\frac{\partial
g_{\mu\gamma}}{\partial x^\sigma}\right)\dot{x}^\sigma\dot{x}^\mu +
g_{\mu\gamma}\ddot{x}^\mu\right],\nonumber
\end{equation}
\noindent Since  $x^a$ are cyclic variables, it reads
the integration constant:
\begin{equation}\label{constant}
g_{cm}\dot{x}^m = N_c,
\end{equation}
\noindent a constant vector. Hence
$
\frac{\mathrm{d}}{\mathrm{d}s}\left(\frac{\partial L}{\partial
\dot{x}^c}\right)=0.
$
Inserting  the terms together, multiplying by $L g^{\mu\gamma}$ and
using (\ref{constant}) the equations of motion are derived:
\begin{equation}\label{motion}
\ddot{x}^\mu + \left\{_{\alpha\beta}^{~\mu}\right\}
\dot{x}^\alpha\dot{x}^\beta =
\frac{1}{2}g_{ab,\gamma}g^{\mu\gamma}N_c g^{ac} N_d g^{bd}.
\end{equation}
\noindent Clearly it is possible to realize that the extra dimensions
induce an external force in the system, that depends on $g_{ab}$ and
$N_c$.
\section{Equation for the visible field}\label{sec:visible}
The main aim now is to compute the gravitational potential in the Newtonian
limit, since galaxies and clusters can be described physically as
Newtonian objects --- corresponding to the approximation in which gravity is
weak. The weak limit is assumed uniquely in the 4-dimensional
spacetime: the deviation $\gamma_{\alpha\beta}$
of the 4-dimensional metric
$g_{\alpha\beta}=\eta_{\alpha\beta}+\gamma_{\alpha\beta}$ is small
 ($\eta_{\alpha\beta}$ denotes the Minkowski metric). For the linearized gravity, the stress (\ref{talfabeta}) is essentially
written from linearized curvature $\mathcal{R}^{(1)}_{\alpha\beta} =
\partial_\mu \left\{_{\alpha\beta}^{~\mu}\right\} -
\partial_\alpha\left\{_{\mu\beta}^{~\mu}\right\}$ as \bege
\mathcal{T}_{\alpha\beta}=-\frac{1}{2}\partial^\mu\partial_\mu\overline{\gamma}_{\alpha\beta}
+ \partial^\mu\partial_{(\beta}\overline{\gamma}_{\alpha)\mu} -
\frac{1}{2}\eta_{\alpha\beta}\partial^\mu\partial^\nu\overline{\gamma}_{\mu\nu}\enge
\noi where $\overline{\gamma}_{\alpha\beta}=\gamma_{\alpha\beta} -
\frac{1}{2}\eta_{\alpha\beta}\gamma$ is the traceless part of ${\gamma}_{\alpha\beta}$ and $\gamma = \gamma_\alpha^{\;\,\alpha}$.
\begin{figure*}
\begin{center}
$\begin{array}{c@{\hspace{0.01in}}c} \multicolumn{1}{l}{\mbox{\bf
(a)}} &
    \multicolumn{1}{l}{\mbox{\bf (b)}} \\ [-0.33cm]
\epsfxsize=3.45in \epsffile{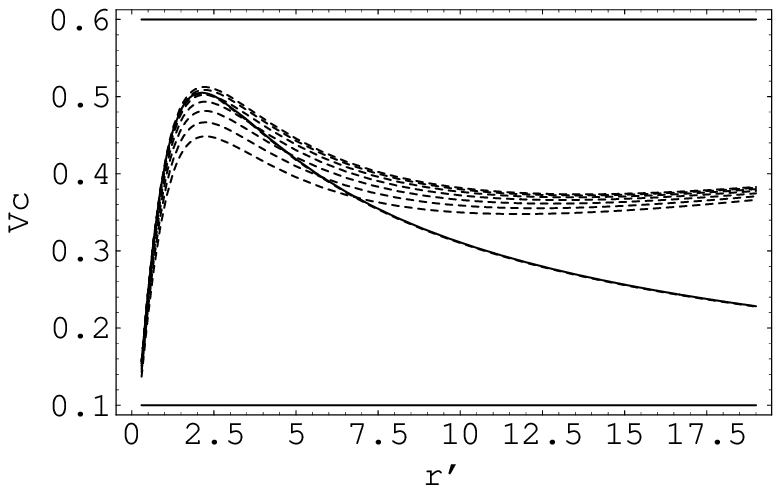} &
    \epsfxsize=3.45in
    \epsffile{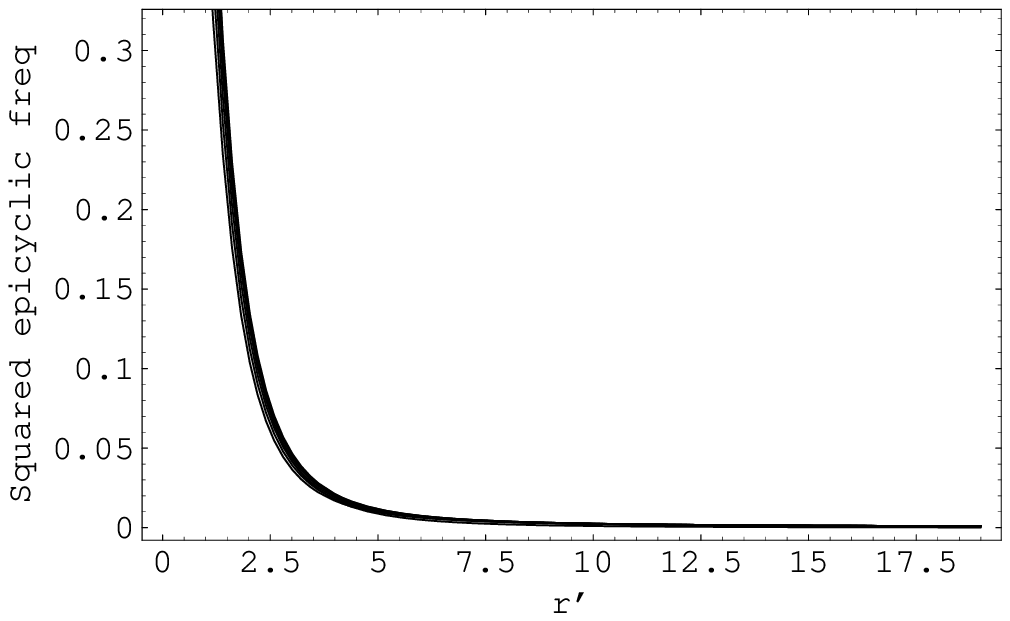} \\ [0.05cm]
\end{array}$
\end{center}
\caption{{\bf (a)} The rotation curves of an effective potential
where the Poisson equation is solved from a Miyamoto-Nagai
\emph{ansatz}, with parameters $a=1$ and $b=0.1$, describing a disk
galaxy, and where the Laplace equation for the extra field is solved
by a Chazy-Curzon disk, with stable cut parameter $c=1.5$. They
correspond to gravitating particles which orbit in a background. The
full line is the curve when there is no extra dimensions, i. e.,
$N_{z_1}=N_{z_2}=0$. The dotted lines represent a set of extra
dimensional solutions for fixed $N_{z_1}=0.1$ and $N_{z_2}$ varying
from $0.8$ to $0.95$, as in \cite{coimbra2}. The parameters $a,b,c$
are chosen in order to the orbital motion be stable. {\bf (b)} The same
set of solutions presented in (a), but now we represent the squared
epicyclic frequency $\kappa^2$. When this last is positive, we
guarantee the stability of the system.}\label{fig2}
\end{figure*}
Linearized gravity has a gauge freedom given by
$\gamma_{\alpha\beta} \mapsto \gamma_{\alpha\beta} +
\pounds_{\xi}\eta_{\alpha\beta}$, where $\pounds_{\xi}$ denotes the Lie
derivative with respect to the generators $\xi^\alpha$ of a differential
diffeomorphism. To the first order, such transformation represents the
same physical transformation  as $\gamma_{\alpha\beta}$. This gauge freedom is used to simplify the linearized Einstein equation. Solving the equation $\partial^\beta\partial_\beta\xi_\alpha =
-\partial^\beta \overline{\gamma}_{\alpha\beta}$ for $\xi_\alpha$, a gauge transformation \cite{wald} that leads to
$\partial^\beta\overline{\gamma}_{\alpha\beta} = 0$  --- similar to the
Lorentz gauge condition --- can be elicited  to obtain the simplified Einstein equation
\bege \label{linear_t}
\mathcal{T}_{\alpha\beta}=-\frac{1}{4}\partial^\mu\partial_\mu
\overline{\gamma}_{\alpha\beta}.\enge \noi
\noi For the extra part it reads \bege \label{linear_extra_t}
\mathfrak{T}_{\alpha\beta} =
\frac{1}{2}\left[\frac{1}{2}(g^{mn}\partial^\mu\partial_\mu
g_{mn})g_{\alpha\beta} - g^{mn}g_{mn,\alpha\beta}\right]. \enge
\noi When gravity is weak, the linear approximation to GR should be valid. There exists a global inertial
coordinate system of $\eta_{\alpha\beta}$ such that \bege
T_{\alpha\beta} = \mathcal{T}_{\alpha\beta} +
\mathfrak{T}_{\alpha\beta} \approx \rho t_\alpha t_\beta, \enge
 \bege\label{super_poisson}  -\frac{1}{4}\partial^\mu\partial_\mu
\overline{\gamma}_{\alpha\beta}+\frac{1}{2}\left[\frac{1}{2}(g^{mn}\partial^\mu\partial_\mu
g_{mn})\eta_{\alpha\beta} - g^{mn}g_{mn,\alpha\beta}\right] = \rho
t_{\alpha} t_{\beta},\enge
\noi where $t_{\alpha}$ is the time direction associated with this coordinate
system. This equation can be interpreted as the modified Poisson
equation considering a universe with more than $3+1$ dimensions.
\section{Extra dimensions and the interpretation of dark
matter}\label{sec:extra}
Define $\overline{\gamma}_{\alpha\beta} \equiv -4\phi$, where
$\phi=\phi(\vec{x})$ is a 3-space scalar field. Furthermore consider a line element $\mathrm{d}s_n^2=\sum_{i=1}^{n}
e^{\psi_i}\mathrm{d}z_i^2$, where
$\mathrm{d}s^2_n$ is the world line for the extra sector, $z_i$ denotes
the extra coordinates and $\psi_i=\psi_i(\vec{x})$ are
potentials associated to extra dimensions.


If one asserts, as a first approximation, that $R_{ab}=0$ (that is
\emph{not} similar to impose vacuum in the extra space, since now $T_{ab}$
becomes $T_{ab}= - \frac{1}{2}(g^{MN}R_{MN})g_{ab})$,
 Eq. (\ref{ricci_explicit_2}) implies at first order the sigma
model $g^{\mu\nu}(\sigma_{,\mu}\sigma^{-1})_{,\nu}=0$ for the extra
part, where $\sigma$ denotes the diagonal matrix representing the metric
associated to the system, which implies
\bege\label{laplace_extra}\partial^\mu\partial_\mu g_{ab} = 0,\enge
\noi yielding the following equation
\bege\label{super_poisson_2}  -\frac{1}{4}\partial^\mu\partial_\mu
\overline{\gamma}_{\alpha\beta}-\frac{1}{2}g^{mn}g_{mn,\alpha\beta}
= \rho t_{\alpha} t_{\beta},\enge \noi or in other words
\bege \label{poisson}\nabla^2 \phi = \rho. \enge
\noi It means that our visible matter density profile is provided
uniquely by the 4-dimensional field. If extra dimensions have some impact on the
theory, certainly it is not as extra matter, but it appears at least
as the term  $\frac{1}{2}g^{mn}g_{mn,\alpha\beta}$
 in (\ref{super_poisson_2}).
In matter models applicable to a galactic context pressure
arises from the velocity dispersion in the motion of particles
or from exchange of momentum among particles through collisions ---
ionized interstellar baryonic gases. An extra pressure for the same
energy-density could be possible for the star gases, which have no
equation of state, but would introduce modifications on the equation
of state of the interstellar gases. Instead, the extra term in
equation (\ref{super_poisson_2}) has the form of a tidal force
produced by the purely geometric effect of the extra dimensions.
Likewise, the equation of motion does not necessarily dismiss the
influence of the extra field, as it can be illustrated by the
following example. A general metric can be written for the
simplified case of six dimensions as $\mathrm{d}s^2 =
-(1-2\phi)\mathrm{d}t^2 + \mathrm{d}\vec{x}.\mathrm{d}\vec{x} +
e^{-\psi}\mathrm{d}z_1^2 + e^{\psi}\mathrm{d}z_2^2$ \cite{coimbra2},
where $\mathrm{d}\vec{x}\cdot\mathrm{d}\vec{x}$ is the
$3$-dimensional line element. Now, in the Newtonian limit the motion
is conceived to be much slower than the speed of light, and
$\dot{x}^\alpha$ can be assumed to be $(1,0,0,0)$ in (\ref{motion}),
as well as the affine parameter $s$ may be approximated by the
coordinate time $t$. Thus it follows that
\begin{equation}\label{motion1}
\frac{\mathrm{d}^2 x^\mu}{\mathrm{d}t^2} +
\left\{_{00}^{~\mu}\right\}=\frac{1}{2}\left\{\left[\frac{\partial
(e^{-\psi})}{\partial
x^\mu}\right]e^{2\psi}N^2_{z_1}+\left[\frac{\partial
(e^{\psi})}{\partial x^\mu}\right]e^{-2\psi}N^2_{z_2}\right\}.
\end{equation}
\noindent As time derivatives of $\phi$ and $\psi$ are neglected,
Eq. (\ref{motion1}) leads to
\beq\label{acceler}
\vec{a} &=& -\nabla \Phi,\\
\label{effective} \Phi&=&\phi +
\frac{1}{2}(e^{-\psi}N^2_{z_{2}}+e^{\psi}N^2_{z_{1}}),\eeq
\noindent that is the equation of motion of test bodies gravitating in an orbit in a given background for our
system with an effective gravitational potential $\Phi$. To find a
form for those potentials, from (\ref{laplace_extra}) and
(\ref{poisson}) it yields
\begin{eqnarray}\label{laplace_equation}
\nabla^2 \psi - \nabla \psi \cdot \nabla \psi &=& 0\\
\label{phi}
\nabla^2 \phi &=& \rho.
\end{eqnarray}
\begin{figure*}
\begin{center}
$\begin{array}{c@{\hspace{0.01in}}c} \multicolumn{1}{l}{\mbox{\bf
(a)}} &
    \multicolumn{1}{l}{\mbox{\bf (b)}} \\ [-0.33cm]
\epsfxsize=3.45in \epsffile{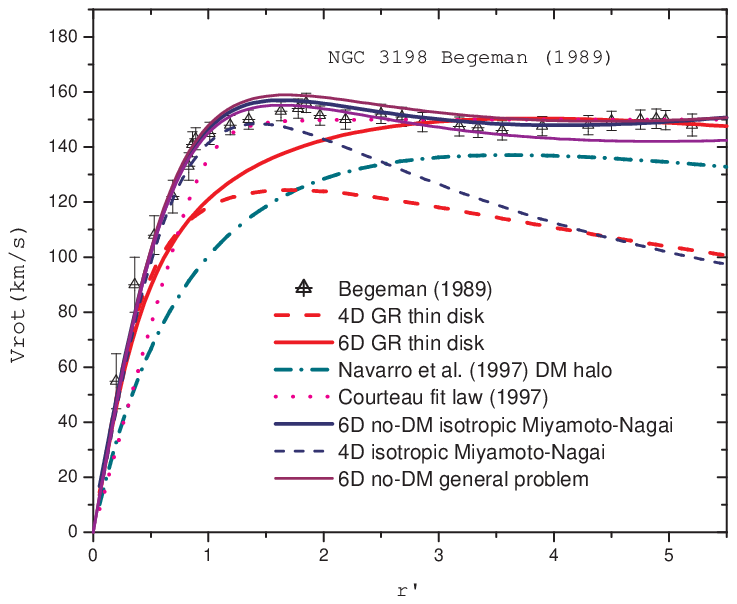} &
    \epsfxsize=3.64in
    \epsffile{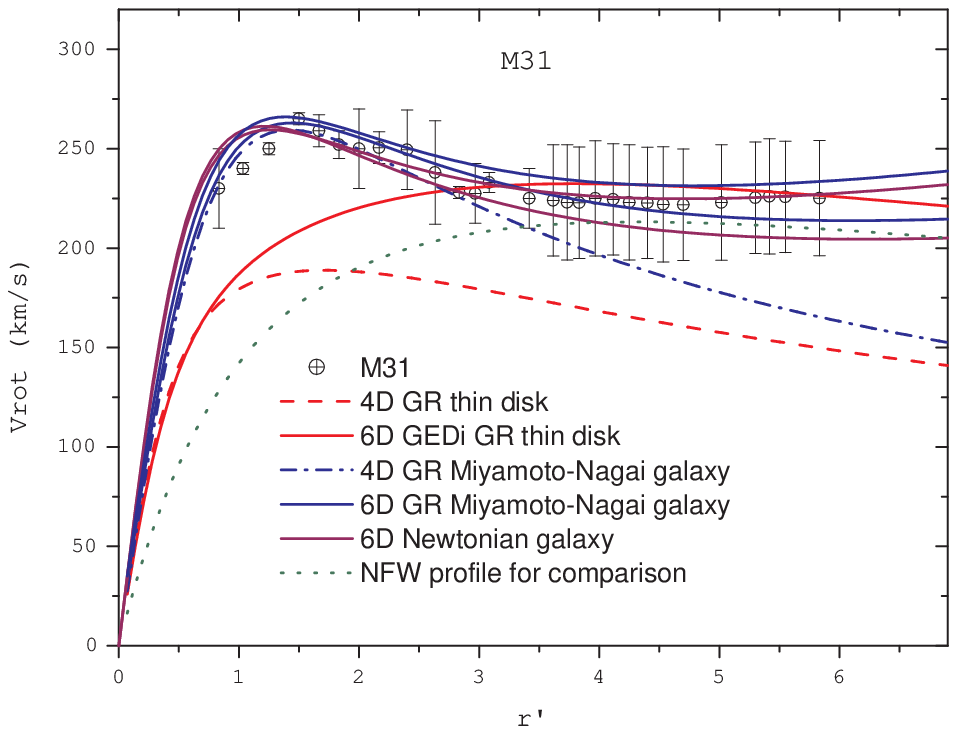} \\ [0.05cm]
\end{array}$
\end{center}
\caption{Values for the general 6-dimensional model (Newtonian
limit): $N_{z_{1}}=0.2$, $N_{z_{2}}=0.8$ and $0.9$, $a=1$, $b=0.01$.
The violet lines are two stable examples of the present model. The
full blue line is the relativistic model presented in
\cite{coimbra3}.}\label{compar1}
\end{figure*}
\begin{figure*}
\begin{center}
$\begin{array}{c@{\hspace{0.01in}}c} \multicolumn{1}{l}{\mbox{\bf
(a)}} &
    \multicolumn{1}{l}{\mbox{\bf (b)}} \\ [-0.33cm]
\epsfxsize=3.65in \epsffile{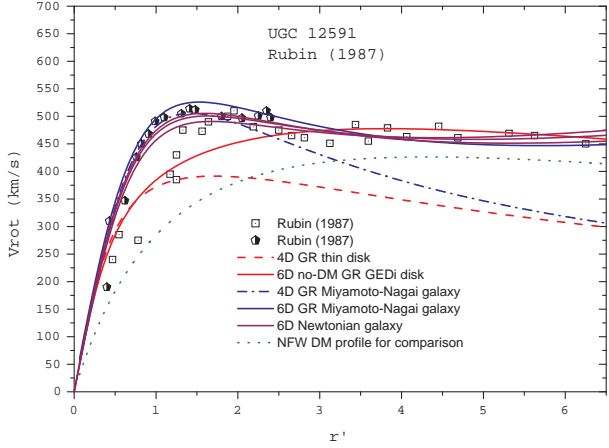} &
    \epsfxsize=3.64in
    \epsffile{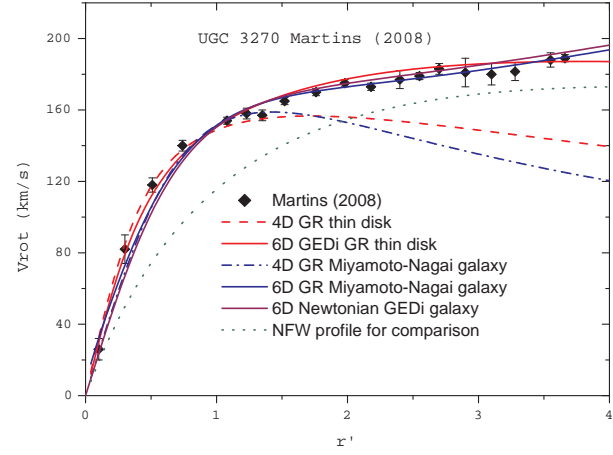} \\ [0.05cm]
\end{array}$
\end{center}
\caption{Values for the general 6-dimensional model:
$N_{z_{1}}=0.18$, $N_{z_{2}}=0.85$ and $0.9$, $a=1$, $b=0.01$. The
violet lines are two stable examples of the present model. The full
blue line is the relativistic model presented in
\cite{coimbra3}.}\label{compar2}
\end{figure*}
\noindent Non-linear terms do not appear, since the $\sigma$
matrix is diagonal. In particular, Eq. (\ref{laplace_equation}) can
be rewritten as
\begin{equation}\label{laplace_2}
\nabla^2 \chi=0,
\end{equation}
\noindent where the identification $
\chi = e^{-\psi}
$ is accomplished, and hence Eq. (\ref{effective}) has the form
\begin{equation}
\Phi = \phi + \frac{1}{2}(\chi N^2_{z_{2}}+ \chi^{-1} N^2_{z_{^1}}).
\end{equation}
\noi Together with Eqs. (\ref{phi}) and (\ref{laplace_2}), it
represents a complete system of equations for a gravitational
effective potential in 4-dimensions, assuming a 6-dimensional universe.
Note as well that (\ref{laplace_2}) is the Laplace equation in flat
3-dimensional space, and therefore $\chi$ can be taken as a
solution of Laplace equation for an appropriate Newtonian source of
any symmetry. The form of $\phi$ can be indeed expressed as the point
source potential $\sim 1/r$ for a Newtonian system.

In a system which acceleration $\vec{a}$, Eq. (\ref{acceler}), has
radial direction, e.g. an idealized system similar to a galaxy, the
rotation curves for a potential $\phi \sim -1/r$ are provided by
\begin{equation}
V_C \sim
\left[\frac{1}{r}+\frac{1}{2}r\psi_{,r}(N^2_{z_{2}}e^{-\psi} -
N^2_{z_{1}}e^{\psi})\right]^{1/2}.
\end{equation}

\section{Solutions}\label{sec:solutions}
Now we look for a way to implement an exact Newtonian solution for
the Poisson equation (\ref{phi}), as well as for the Laplace
equation (\ref{laplace_2}). As  a system similar to a galaxy is
regarded, the Poisson equation can be implemented e. g. by a
Miyamoto-Nagai potential, and for the case of Laplace equation it
can be used a Chazy-Curzon  solution for an axisymmetric
configuration. Namely the Miyamoto-Nagai \emph{ansatz} is written as
\cite{nagai} \bege\label{miyamoto}
\phi(R,z)=-\frac{M}{\sqrt{R^2+(a+\sqrt{z^2+b^2})^2}}, \enge \noi
where $a, b$ are positive constants. The Chazy-Curzon solution for a
particle of mass $m$ in the position $z=z_0$ is given by
\cite{chazy,curzon} \bege \chi = \frac{2m}{R}, \enge where
$R=\sqrt{r^2+(z-z_0)^2}$. In this last, it is possible to fix
$z_0=0$, and to introduce the cut method to generate a disk solution
\cite{binney}, such that $z\mapsto |z|+c$, where $c$ is the cut
parameter. Fixing values for $a$, $b$ and $c$ in such a way that
$b/a \sim 0.1$ -- providing a similar light distribution of a disk
galaxy \cite{binney} --  and $c>1$ (by stability issues and to
prevent relativistic disks \cite{coimbra2}) it is possible to find
the set of curves given in Fig. \ref{fig2}a.

The stability of this last configuration is guaranteed by the
positive epicyclic Newtonian parameter $\kappa^2=\partial^2
\Phi/\partial r^2 + 3V_C^2/r^2$, as shown in Fig. (\ref{fig2}b).

A more useful manner to analyze the effective potential is writing it as
\bege\label{effective_2} \Phi = \phi + C \cosh (\psi + \delta),\enge
\noi where $C$ and $\delta$ are constants to be determined. This is the  correspondent Newtonian potential in the present formalism. Clearly,
from Eq. (\ref{effective}) those constants are related to
$N_{z_{1}}$ and $N_{z_{2}}$ as
\beq C^2 &=& \frac{N_{z_{2}}^2 - N_{z_{1}}^2}{4},\qquad
\tanh \delta = \frac{N_{z_{1}}^2}{N_{z_{2}}^2}.\nonumber \eeq\noi Given the present results, one can ask about the nature of the potential (\ref{effective_2}). The term  comes exclusively from
extra dimensions and  it may be related to a ``dark matter'' potential as
\bege \phi_{{\rm extraD}} = \phi_{DM} = C \cosh (\psi + \delta),\enge
\noi showing that our approach is equivalent to a dark matter
effect, although the concept above is completely apart from the
concept of extra matter.
\section{Probing results with real galaxies}\label{sec:galaxies}
Taking into account only the stable curves, in Figs. \ref{compar1}
and \ref{compar2} we compare with some optically observed rotation
curves of spiral galaxies.

Here we are not providing a composition of halo dark matter velocities
plus the disk gas and the velocities of stars. What is happening is
that the clean stable calculated curves are simply fitting the
region of interest. The not surprising {\it ad hoc} adjustment of
$N_{z_{1}}$ and $N_{z_{2}}$ actually could assert nothing about the
astrophysical role of the extra dimensions in the model. However, the
calculation of stable configurations brings realizable
values for $N_{z_{1}}$ and $N_{z_{2}}$, which makes it possible to
visualize a minimum representation of a real disk galaxy. We furthermore
compare to some phenomenological models used in astrophysics as
Navarro, Frenk \& White \cite{navarro}. The observed data is taken
from \cite{sofue,rubin,begeman}, but for an alternative data source
see \cite{rotation}, where our model is also successful.


Note that our results reproduce with great fidelity the shape of
non-planar curves, as that appearing in Fig. \ref{compar2}(b) (data
taken from \cite{martins}).

\section{Concluding remarks}\label{sec:conclude}
At the present work we showed how to calculate a ``dark''
potential from extra-dimensional imprints. If extra dimensions have
some impact on the theory, certainly it is not as extra matter, but
it appears at least as tidal force produced by the purely geometric
effect of the extra dimensions, namely the
$\frac{1}{2}g^{mn}g_{mn,\alpha\beta}$ term in
(\ref{super_poisson_2}). This is confirmed by the extra force
appearing in the equations of motion.

The Newtonian limit is calculated and we accomplished the visible
potential that has two main components: a term from 4-dimensional
metric and another term  coming exclusively from extra-dimensional effects. The
4-dimensional term obeys the Poisson equation and the extra term is
solved from a Laplace equation. Two situations are provided: a first
approximated model and a second that arises from exact solutions of
Laplace and Poisson equation, respectively the Chazy-Curzon disk
solution and Miyamoto-Nagai \emph{ansatz}. Those examples are
calculated for a case where two extra dimensions are taken into
account. Moreover, any analogous weak field potential decomposition,
as the effective ones for elliptic galaxies or globular clusters, in
which $\phi$ would be a different potential with the appropriate
symmetry, holds for this formalism.

The extra field arises naturally from metric extra terms, and the
modification of gravity shown from the calculation of rotation
curves, can present an explanation about the missing mass effect in
astronomy. The effective potential calculated here is a source to
explain the phenomenological results found in \cite{coimbra3} and
\cite{coimbra2}. The extra part of this effective potential has the
form $C \cosh [\psi(x^\alpha) + \delta]$. The gradient of such
potential represents the extra force associated to the presence of
extra dimensions. It modifies the rotation curve profiles and rise
anomalies in potential associated to clusters. The Laplacian
corresponds to a ``dark`'' density profile, that is negligible
compared to the visible density $\nabla^2 \phi = \rho$ (see Section
\ref{sec:extra}). Furthermore, the extra-dimensional proposal is
consistent with a a more general length scale and galactic
morphology.

The  present calculation is an alternative approach to understand
the dark matter problem. In fact, to solve the dark matter problem
means to solve a long list of cosmological anomalies. In this
aspect, as a complement, considerations about cluster lensing are
also addressed in \cite{coimbra1}.

\acknowledgments
The authors are extremely grateful to Patricio S. Letelier (\emph{in
memoriam}) for the important and profound contribution concerning
interpretation and discussion of the present results. The work of
C.H. C.-A. is supported by Funda\c c\~ao Arauc\'aria and PDEE/CAPES
Programme under Grant No. 3874-07-9. R. da Rocha is grateful to CNPq 472903/2008-0 and 304862/2009-6 for financial support. \\

\end{document}